\documentclass[twocolumn,aps,prb,showpacs]{revtex4}
\usepackage{amsmath} 
\usepackage{amssymb} 
\usepackage{graphicx} 
\usepackage{epsfig} 
\usepackage{color}
 \usepackage{rotating}
 
\begin{document} 
\title{Unconventional metallic conduction in two-dimensional Hubbard-Wigner lattices} 
\author{S. Fratini$^{1,2}$ and J. Merino$^3$} 
\affiliation{$^1$ Institut N\'eel-CNRS and Universit\'e Joseph Fourier,
  Bo\^ite Postale 166, F-38042 Grenoble Cedex 9, France \\ 
$^2$ Instituto de Ciencia de Materiales de Madrid, CSIC, Sor
  Juana In\'es de la Cruz 3, E-28049 Madrid, Spain\\
$^3$ Departamento de F\'\i sica Te\'orica de la Materia Condensada, Universidad Aut\'onoma de Madrid, Madrid 28049, Spain}

\begin{abstract}
The interplay between long-range and local Coulomb repulsion 
in strongly interacting electron systems is explored through a two-dimensional
Hubbard-Wigner model. An unconventional metallic state is found in which 
collective low-energy excitations characteristic of the Wigner crystal induce 
a flow of  electrical current despite the absence 
of one-electron spectral weight at the Fermi surface. Photoemission
experiments on certain quarter-filled layered molecular crystals should observe
a gap in the excitation spectrum whereas
optical spectroscopy should find a finite Drude weight indicating metallic 
behavior.         
\end{abstract}
\date{\today} 
\maketitle

\section{Introduction}

Coulomb interactions in two dimensions (2D) can lead to different forms of
electron localization ranging from Mott insulators\cite{Imada1} to Wigner
crystals \cite{Wigner}. Half-filled narrow-band systems
such as cuprate \cite{Bonn,Lee}
and organic superconductors \cite{Ishiguro,McKenzie1} display
insulating states in their phase diagrams which are not described by band theory
approaches. This is due to the strong local Coulomb repulsion
driving the system to a Mott metal-insulator transition (MIT)
which localizes electrons at each site of the lattice.
On the other hand, a gas of electrons in a positive
uniform background can localize through Wigner crystallization when the
{\it long-range} Coulomb repulsion overcomes the kinetic energy, which
is relevant to the MIT observed in two-dimensional electron gases
(2DEG) at sufficiently  low densities
 \cite{Kravchenko,Abrahams}. Besides, there are many 2D crystals
with non-integer filled narrow bands which
are neither Mott nor Wigner insulators but rather display "Wigner
crystallization on an underlying lattice" (WL) \cite{Hubbard78},
with both aspects of electron localization.
The family of quarter-filled layered organic materials,
$\theta$-(BEDT-TTF)$_2$MM'(SCN)$_4$ (M=Rb, Cs, Tl,M'=Zn, Co)
represent clean realizations of WL displaying a subtle 
competition between charge ordering,
superconductivity and unconventional metallic phases in their phase
diagram \cite{Seo}.  
These systems pose the challenging question of whether the interplay between
short (Mott) and long-range Coulomb interactions (Wigner)  \cite{Camjayi08,Pankov} {\it on a lattice} 
can lead to new ground states and excitations and whether such novel states of matter
can be experimentally tested.

Extensions of the Hubbard model including the long-range Coulomb 
interaction have mostly been
limited to one-dimensional Wigner lattices
 \cite{Hubbard78,Valenzuela03,Fratini04,Horsch06,Daghofer07}.
In these systems the degeneracy of the classical ground state at
non-integer fillings leads, in the strongly interacting regime, to 
low-energy excitations consisting of domain walls with fractional charge. 
This type of collective excitations, which 
play a major role in the melting of the Wigner
lattice on the one-dimensional chain \cite{Fratini04,Horsch06}, 
have been much less explored in two dimensions. \cite{Tsiper,Pollmann}
Here we find that collective excitations in 2D can give rise to 
metallic conduction in a charge ordered  metallic state (COM)
even though there is a finite charge gap in the one-electron spectrum. 
By combining conductivity and photoemission/tunneling experiments, 
such an anomalous metallic phase could be clearly  
identified and discerned from a more conventional charge density wave
(CDW) metal, 
in which the conducting behaviour is ensured by portions of the
Fermi surface that remain gapless.

The paper is organized as follows. The model and the numerical method 
that we use are presented in Sec. II. The main
results  are contained
in Sec. III, and their possible relevance to experimental systems is
discussed in Sec. IV.
In the appendices we introduce an
analytical model valid in the limit of strong long-range Coulomb
interactions, that provides a clear interpretation of the numerical
results in terms of defects of the checkerboard ordering.

\section{Model and method}

The minimal model to describe the effect of the long-range part of the Coulomb
interaction in 2D is 
the Hubbard-Wigner model (HWM) at one-quarter filling on a square lattice:
\begin{equation}
H=-t\sum_{<ij>\sigma}(c^\dagger_{i\sigma} c_{j\sigma} +c^\dagger_{j\sigma}c_{i\sigma}) + U \sum_i n_{i\uparrow}n_{i\downarrow}
+\sum_{ij}V_{ij}n_in_j,
\end{equation}
where $c^\dagger_{i\sigma}$ creates an electron with spin $\sigma$ on
site $i$,  
the kinetic energy is parametrized by the  
hopping amplitude $t$ between neighboring sites and 
the onsite Coulomb interaction by $U$. The long-range contribution is: 
$V_{ij}=V/|i-j|$, with $|i-j|$ the 
distance between two different sites on the lattice and $V$ 
a parameter controlling the strength of the non-local 
Coulomb repulsion. The model is solved using Lanczos diagonalization
on finite clusters of up to $L=18$ sites with periodic boundary conditions. 
The Coulomb potential between two different sites inside the cluster
is calculated from Ewald summations
which account for an  infinite periodic array of simulation cells.
In order to single out the effects introduced
by the long-range part of the interaction we compare results of the full
model with the extended Hubbard model (EHM) 
containing $U$ and a nearest-neighbor $V$ only. 
We deliberately choose a large
$U=100 t$  in order to disentangle the relevant low-energy excitations
of the system,  focusing primarily on the charge sector. 
Nevertheless, the main conclusions obtained here remain valid 
as long as $U$
is larger than the bandwidth, $8t$, which includes  
the realistic case $U\approx 20t$ relevant to the 
organic cystals: $\theta$-(BEDT-TTF)$_2$X.

\section{Results}
\subsection{Phase diagram}

\begin{figure}
\centering
\epsfig{file=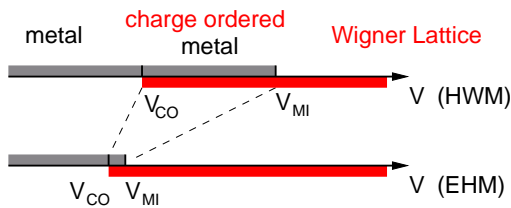,width=7.cm,angle=0,clip=}\\
\caption{Sketched phase diagram of the 
quarter-filled Hubbard-Wigner model. The uniform metallic phase is 
stabilized by the inclusion
of long-range Coulomb interactions (HWM) compared to the model with short range
interactions (EHM) and an intermediate charge ordered
metallic phase emerges in between the uniform metal and the Wigner lattice.
}
\label{fig:PD}
\end{figure}

\begin{figure}[t]
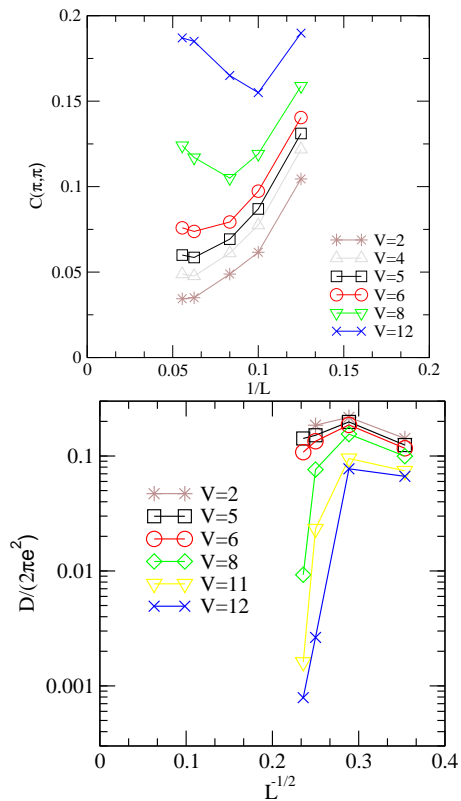

\centering
\epsfig{file=Fig2a.eps,width=5.5cm,angle=0,clip=}
\epsfig{file=Fig2b.eps,width=6cm,angle=0,clip=}
\caption{Finite-size scaling of the static charge correlation
function evaluated at $(\pi,\pi)$ (a) and Drude weight (b) for different
 values of $V$. The charge ordering transition is estimated to occur at $V_{CO}\approx 4t$
from the linear extrapolation  to the
thermodynamic limit of $C(\pi,\pi)$ vs. $1/L$. 
The Drude weight is plotted as a function of $1/\sqrt{L}$,
displaying insulating behavior for $V>V_{MI} \approx 8t$.}
\label{fig:scaling}
\end{figure}

The phase diagram of the HWM obtained from exact
diagonalization (ED) of finite-size clusters is illustrated 
in Fig. \ref{fig:PD},  
where two distinct phase transitions can be
identified. The first one corresponds to the charge ordering instability 
of the homogeneous metal, 
indicated by the presence of non-zero charge correlations at the critical
wavevector ${\bf Q}=(\pi,\pi)$. 
In Fig. \ref{fig:scaling} (a) we show the finite size scaling 
of the charge correlation function on $L=8, 10, 16, 18$ clusters as a
function of $1/L$.  The extrapolation to the thermodynamic
limit suggests a charge ordering transition at about $V_{CO}\approx (4-5)t$
in good agreement with previous results using the Path Integral Renormalization
Group (PIRG) on $4\times4$, $8\times8$ and $12\times12$ clusters
(see Fig. \ref{fig:scaling} and  Ref. \onlinecite{Imada2}).
This value
is much larger than the 
$1.5t$ obtained for the EHM (see also Ref. \onlinecite{Calandra02}). This indicates 
that the uniform metallic phase is notably \textit{stabilized} by the
inclusion of long-range interactions \cite{Tsiper,Imada2}.
To trace back the physical origin of this phenomenon, we observe that
the CO instability can be understood through the divergence of the  
charge susceptibility in the Random Phase Approximation (RPA): $\chi_c({\bf Q})=\chi_0({\bf Q})/\left [ 1+V({\bf
  Q})\chi_0({\bf Q})\right ]$ 
with $\chi_0({\bf Q})$ the bare susceptibility independent of the
Coulomb repulsion and $V({\bf Q})$ the interaction potential in 
reciprocal space at wavevector ${\bf Q}$. 
Since the absolute value of 
$V({\bf Q})$ in the model with long-range interactions is much reduced
(by a factor of $2.5$) 
as compared to the case with nearest-neighbor interactions alone,
the above argument predicts  $V_{CO}\simeq 4t$ in the HWM, in 
 agreement with the ED result. Remarkably, the three different estimates
based on ED, PIRG and the RPA scaling argument lead to the same
value of $V_{CO}$. This result obtained by
considering the full long-range Coulomb potential
already settles a long-standing puzzle in quarter-filled
layered organic conductors, many of which are found to be metallic even for
values of $V$ comparable or larger than the bandwidth of the material.

The second critical point corresponds to the metal-insulator
transition, signaled by the vanishing of the Drude weight. The finite-size scaling 
of this quantity is shown in Fig. \ref{fig:scaling} (b).   
The Drude weight displays a large drop at about $V_{MI}\approx (7-8)t$ after
extrapolating to the thermodynamic limit. 
The broad region $V_{CO}<V<V_{MI}$ therefore corresponds to a {\em charge ordered,
metallic phase}. This phase, which can be easily discerned from
a metallic CDW  (see below), might be  intimately 
related to the hybrid phase predicted at the quantum melting  of the 
Wigner crystal in the continuum. \cite{Spivak,Waintal} 
The  robustness of the COM phase can indeed
 be contrasted to the model with short range interactions, where 
a rather narrow  intermediate region (if any) arises: 
between $V_{CO}\simeq 1.5t$, and $V_{MI}\simeq 1.8t$ also consistent 
with the COM found with cluster dynamical mean-field theory \cite{CDMFT}.
Spin correlations are also enhanced  
in the same parameter range, suggesting that charge and spin do order
together in the  metallic phase before the metal-insulator transition
occurs. Such spin order corresponds to antiferromagnetism between 
charge rich sites due to ``ring'' exchange processes \cite{McKenzie2}. 
It can be noted that using  a more realistic value of $U=20t$
significantly increases the value of $V_{MI}$, while leaving 
$V_{CO}$ essentially unchanged, which effectively broadens 
the region of stability of the COM phase.

\subsection{Excitation spectra}

\begin{figure}
\epsfig{file=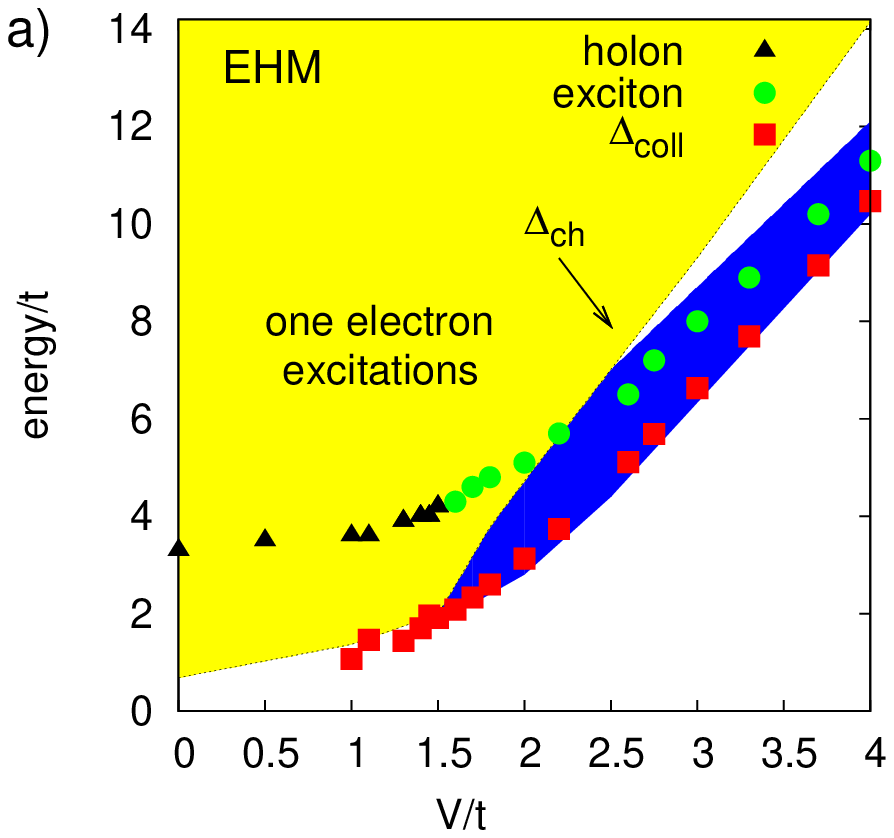,width=8.5cm,height=6cm,angle=0,clip=}
\epsfig{file=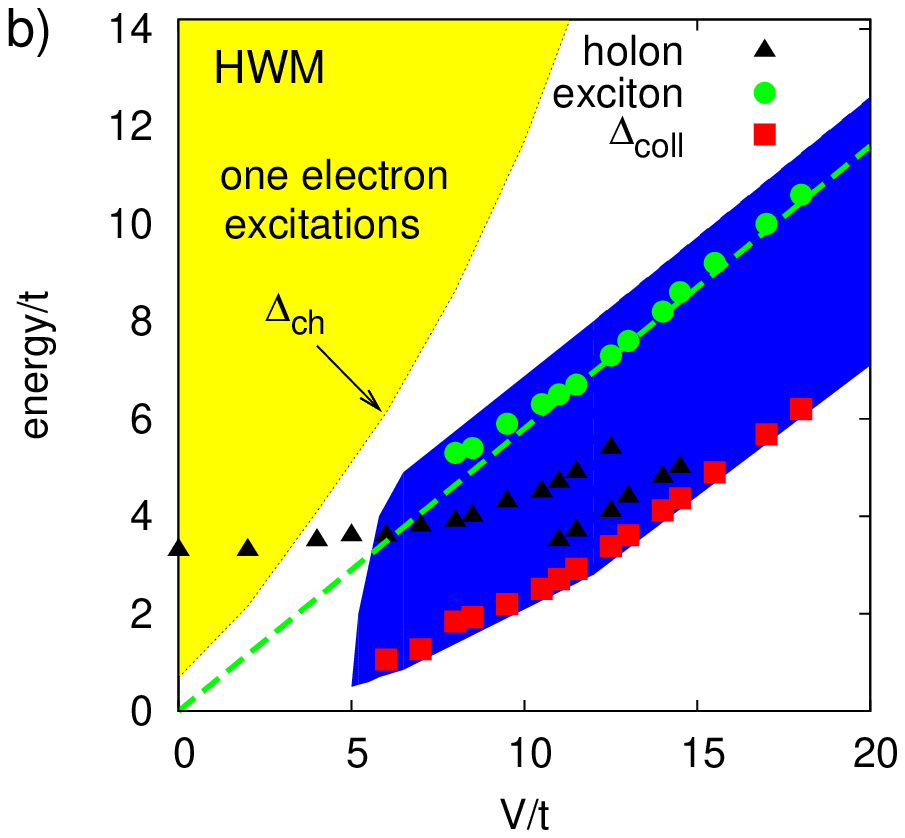,width=8.5cm,height=6cm,angle=0,clip=}
\caption{(Color online) Excitation spectra for the models  
with (a) nearest neighbor (EHM) and (b) long-range interactions (HWM)
on $L=16$ sites. 
     The symbols represent respectively the
    holon (triangles) and exciton (dots) peaks in the optical
    conductivity, and the lowest excitation of finite energy as given
    by the ED  (squares). 
    The light shaded (yellow) region is
delimited by the single-particle  charge gap $\Delta_{ch}$, while the dark 
    (blue) region corresponds to the collective excitations of the
    Wigner lattice, bounded from below by $\Delta_{coll}$. 
The dashed 
    line is the predicted exciton energy  in the thermodynamic limit, 
    $0.62V$ (see the appendix). 
}\label{fig:gaps}
\end{figure}

As we proceed to show, the charge
ordered metallic phase found here shares common aspects
of both  Mott and Wigner physics. 
Its anomalous properties are a direct manifestation of
the collective nature of its low-lying
excitations, originating from the long-range Coulomb repulsion
as well as the strong correlation effects associated
with the on-site Coulomb interaction in a narrow-band system. 
To characterize the excitation spectrum 
we now introduce two representative quantities: the
single-particle charge gap $\Delta_{ch}$ measured
in photoemission experiments, that represents the
energy required to add/remove an electron to/from the system;
the optical gap $\Delta_{opt}$ defined as the the lowest charge excitation
allowed by the dipolar matrix element, corresponding to 
the onset of finite-frequency absorption in an optical experiment. 
The former carries
information on the one-particle excitations, the latter
gives access to the local charge fluctuations, pertaining to the 
collective sector of the spectrum. While 
such quantities coincide for non-interacting electrons, their comparison 
in an interacting system provides valuable information on the excitation
spectrum. These quantities are
plotted in Fig. \ref{fig:gaps}(a) and (b) 
respectively for the EHM and HWM in an $L=16$ cluster, and allow us 
to analyze the relative role played by the single
particle against the collective excitations.

\begin{figure}
\includegraphics[width=8.5cm]{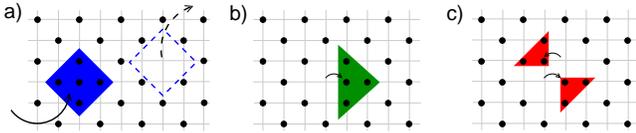}
\caption{(Color online) Lowest-energy excitations 
of the checkerboard Wigner lattice: (a) Charged  (monopoles),
  (b) neutral (dipolar)  and (c) (quadrupolar) defects. (a)
  corresponds to the process of adding/removing a particle to form 
an interstitial/vacancy, which determines the charge gap
$\Delta_{ch}=1.62V$; (b) is a local (dipolar) charge
fluctuation of energy $\Delta_{opt}=0.62 V$, that
 can be viewed as a tightly bound 
interstitial-vacancy pair, or an exciton; 
(c) is the next
(quadrupolar) excitation of energy $\Delta_{quad}=0.65V$.  
\label{fig:defects}
}
\end{figure}

Let us consider
the EHM first. At large $V\gg t$ the charges order in the 
checkerboard pattern shown in Fig. \ref{fig:defects}.  The process of adding 
an electron or a hole (Fig. \ref{fig:defects}a)  
defines the one-particle charge gap $\Delta_{ch}=4V$. 
 Reducing the interaction strength 
leads to a progressive closing of $\Delta_{ch}$ 
(the boundary of the yellow region in Fig. \ref{fig:gaps}a), 
that eventually triggers the metallization transition 
and the consequent melting of the charge ordered state. 
In addition to the usual one-particle excitations, 
the motion of a particle in the EHM from the perfect
checkerboard pattern to one of its
neighboring unoccupied sites gives rise to 
a local (neutral) charge fluctuation depicted in 
Fig. \ref{fig:defects}b. The latter costs an energy 
$\Delta_{opt}=3V$, which at large $V$ lies below the single-particle
gap. As $V/t$ decreases, however, 
the quasiparticle excitations rapidly 
take over owing to their large kinetic energy gain $\sim 8t$.
As can be seen  in Fig. \ref{fig:gaps}a,  
in the region around $V_{MI}$ there is no longer a clear-cut 
distinction between the one-electron and the collective sectors, and
the neutral charge fluctuations do not play a major role in  the
transition mechanism. 
In fact, the critical point  in the case of short range interactions
can be estimated by comparing $\Delta_{ch}$
with the bandwidth $\approx 8t$: this qualitative argument 
yields $V_{MI}\lesssim
2t$ (the $\lesssim$ sign accounts for the bandwidth reduction 
due to the local  $U$), in agreement with the numerical result 
$V_{MI}\simeq 1.8 t$.

This picture changes drastically when the full long-range Coulomb 
repulsion is included, as in the Hubbard-Wigner model Eq. (1). 
The key point is that 
in this case  {\it the defects of the checkerboard pattern 
are no longer confined}. \cite{Andreev} 
Since the electrostatic energy is essentially insensitive to local 
details, being mostly determined by the 
long-range tails of the Coulomb potential,
several electronic configurations can be found which are
almost degenerate with the local defect of Fig. \ref{fig:defects}b.
For example, the energy of the quadrupolar excitation shown in  
Fig. \ref{fig:defects}c is only $0.03V$ larger than that of
Fig. \ref{fig:defects}b, making these two states  practically 
degenerate for any realistic value of $V$.  
As a result,  the electron can resonate between these states,
forming a ``charge droplet'' that can propagate coherently at long 
distances with a net kinetic energy gain $\propto t$.
This phenomenon is directly reflected in the excitation spectrum shown
in Fig. \ref{fig:gaps}b: a collective, itinerant  
defect state  emerges
well below both the optical gap $\Delta_{opt}$ and the charge gap 
$\Delta_{ch}$, whose energy $\Delta_{coll}\simeq\Delta_{opt}- 4.2 t$
qualitatively agrees with the prediction of the defect model,
$\Delta_{coll}\simeq\Delta_{opt}- 2\sqrt{2}t$ derived in the appendix.

\begin{figure}[t]
\centering
\includegraphics[width=5.5cm]{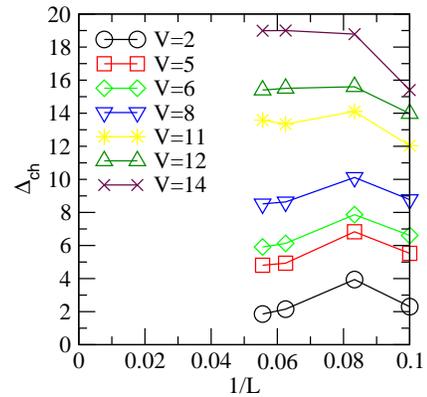}
\caption{ Finite-size scaling of the single-particle charge gap, $\Delta_{ch}$,
for different cluster sizes. Linear extrapolation to the thermodynamic limit indicates
that the charge gap opens at $V \approx 4t\approx V_{CO}<V_{MI}$ suggesting a
phase with a finite charge gap, charge order and non-zero Drude weight (see also Fig.
\ref{fig:scaling}).}
\label{fig:chargegap}
\end{figure}

We see that due to the delocalization of defects in the presence of
long-range interactions, these collective excitations can gain 
a kinetic energy comparable to the one-particle excitations.
As a result, 
the separation of energy scales between the 
collective sector and the single-particle sector, characteristic of
the strongly interacting limit,  survives 
down to $V_{MI}$ and below. 
Accordingly,   the metallization transition 
as well as the resulting charge ordered metallic phase in the region
$V_{CO}<V<V_{MI}$ appear to be entirely driven by the low-energy collective
sector \cite{Tsiper}.
The finite-size scaling shown in Fig. \ref{fig:chargegap} 
strongly suggests that the one-particle charge gap $\Delta_{ch}$  
remains finite within the metallic phase: the charge gap 
is found to open concomitantly with 
the charge ordering transition occurring at $V_{CO} \approx 4t <V_{MI}$,
{\it i. e.} in the range of parameters where the system has a non-zero
Drude weight. The region $V_{CO}<V<V_{MI}$ therefore corresponds to a
charge ordered, metallic phase, with a vanishing single-particle
weight at low energy.

\subsection{Spectral probes}

The  excitation spectrum described in the preceding paragraphs
provides a way to univocally  characterize  
the anomalous 
COM phase predicted here.
First of all, in conjunction with the charge ordering observed by X-ray
diffraction or NMR,
photoemission or tunneling experiments should find
a finite  single-particle gap (see Figs. \ref{fig:gaps}b and
\ref{fig:chargegap}),  
despite a markedly metallic behavior.  In fact both 
the large value of the charge gap as well as  
the energy dispersion of the lowest-lying single-particle
excitations are well captured by the defect model presented in the appendix. 
There it is shown that the dispersion of the charged interstitials of
Fig. \ref{fig:defects}a is accurately described by
the following formula:
\begin{eqnarray}
  \label{eq:disptext}
E_k &=& E_0-2 t_1 [\cos (2 k_x) + \cos (2 k_y)] \\
\nonumber & & -
2 t_2 [\cos (k_x + k_y) + \cos(k_x - k_y)].
\end{eqnarray}
representing the motion of interstitials respectively 
along the principal axes and
along the diagonals of the square lattice.
The physical picture that emerges from the dispersion relation
Eq.(\ref{eq:disptext}) is that of interstitial 
defects moving as a separate fluid {\it
on top} of the charge ordered pattern, which bears a strong resemblance
with the hybrid phase of the Wigner crystal in the
continuum.\cite{Waintal}
An important implication of
the defect model is that the  hopping parameters
$t_1,t_2$ governing the motion of interstitials
are essentially constant throughout
the charge ordered phase: the one-electron
band dispersion depends on the interaction strength $V$ only via the
rigid energy shift $E_0$, related to the charge gap $\Delta_{ch}$. This
can once again be contrasted with the standard CDW picture where the
band dispersion shrinks as $\propto t^2/V$ upon increasing $V$.

\begin{figure}[t!]
  \centering
\vspace{.5cm}
\epsfig{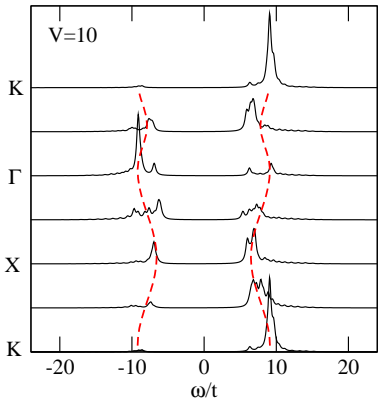}
\caption{
 The spectral function $A(k,\omega)$
   as obtained from Lanczos
  diagonalisation of the Hubbard-Wigner model on a $L=16$ cluster
  at $V/t=10$. The different points in the Brilloun zone are
  $\Gamma=(0,0)$, $X=(\pi,0)$, $K=(\pi,\pi)$. The red dashed line is
  the band dispersion Eq. (\ref{eq:disptext}) of defects moving along 
 the diagonals of the square lattice. 
 The offset energy $E_0$  has been adjusted 
to fit the  overall position of the bands. Note that the critical
values $V_{CO}$ and $V_{MI}$ at this system size are 
larger than those extrapolated to the thermodynamic limit, so that
the value  $V/t=10$ is representative of the charge ordered metallic region
$V_{CO}<V<V_{MI}$.}
\vspace{.5cm}
  \label{fig:spectral}
\end{figure}

Fig.\ref{fig:spectral} shows the  spectral function $A(k,\omega)$ obtained
from the exact diagonalisation of a $L=16$ cluster for
$V=10t$, which lies well within the charge ordered metallic phase for
this system size.  The main peak  
clearly follows the dispersion of the defect band 
[Eq. (\ref{eq:disptext}), with $t_1=0$, $t_2= -0.3 t$]
representing the motion of interstitials along the diagonals of the square
lattice.  A similar agreement applies throughout the charge ordered
metallic phase $V_{CO}<V<V_{MI}$, improving at larger values of $V$.

\begin{figure}
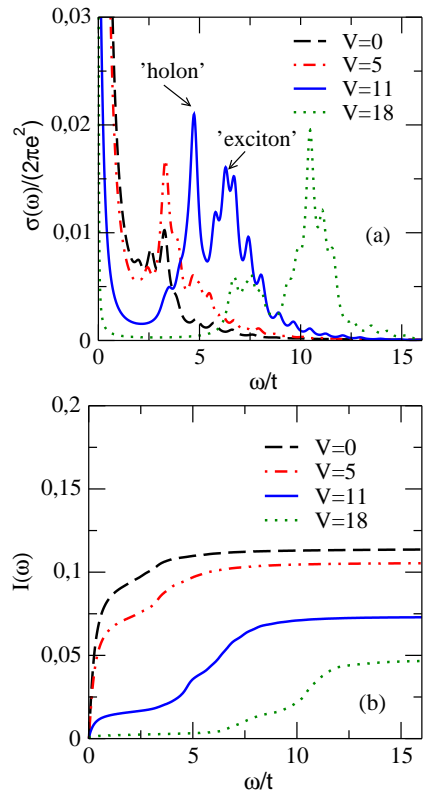

\begin{center}
\epsfig{file=Fig7a.eps,width=5.5cm,angle=0,clip=}
\epsfig{file=Fig7b.eps,width=5.5cm,angle=0,clip=}
\end{center}
\caption{(Color online) Optical properties
across the CO transition in the Hubbard-Wigner model.
(a) The 'holon' absorption band of the strongly correlated 
metal is strongly enhanced by $V$, dominating the optical 
conductivity spectra for 
$V<V_{CO}$.
Within the charge ordered metallic phase, $V_{CO}<V<V_{MI}$ a broader band at 
about $0.62V$ arises due to bound excitons
(Fig. \ref{fig:defects}b). (b) Integrated spectral weight of the 
optical conductivity for the same parameters as in (a). A strong 
transfer of spectral weight from low energies to high energies across
the CO transition is evident. Results are shown for a $L=16$ cluster.}
\label{fig:opt}
\end{figure}

The calculated optical conductivity also reveals clear signatures
of the predicted COM phase. As shown  in Fig. \ref{fig:opt}(a), 
in addition to a Drude peak (presumably anomalous, being  
possibly carried by collective excitations),
the optical spectra exhibit two distinct  finite-frequency absorption bands, 
confirming the 'mixed' nature of such intermediate phase: 
a low-frequency 'holon' peak at about $(3 - 4)t$ 
originating from the short-range correlations, and enhanced by the  non-local Coulomb
repulsion \cite{Greco,Dressel}, plus a broader excitonic band at $\Delta_{opt}=0.62V$,
corresponding to the dipolar defects of the Wigner lattice. 
Fig.  \ref{fig:opt}(b) shows the integrated optical spectra, $I(\omega)$,  
across the CO transition. For $V=0$, most of the spectral weight is
in the Drude weight and $I(\omega)$ saturates rapidly with $\omega$. 
As $V$ increases a strong redistribution of spectral weight 
occurs with the low energy optical weight being transfered   
to higher energies (of the order of $\Delta_{opt}$). The total integrated
spectrum is a measure of the kinetic energy of the system as
derived from the $f$-sum rule\cite{Maldague}. Fig.  \ref{fig:opt}(b)
shows how the electron kinetic energy is suppressed by $V$
due to electron correlation effects close to the CO transition.

\section{Concluding remarks}

The quarter-filled layered organic crystals of the $\theta$-type are
ideal realizations of a Wigner lattice in which some of the existing 
experiments could be interpreted in light of the present scenario.
The horizontal charge ordered pattern observed through X-rays \cite{Mori07} 
below the CO temperature $T_{CO}=190$K in $\theta$-(ET)$_2$RbZn(SCN)$_4$, 
is consistent with CO driven by the off-site Coulomb repulsion but inconsistent
with any Fermi surface nesting vector. In the metallic salt
$\theta$-(ET)$_2$CsZn(SCN)$_4$, both stripe-type and three-fold 
CO patterns are observed, both related to strong 
off-site Coulomb repulsion \cite{Mori07}. Metallic phases above
$T_{CO}$ display 'bad' metallic behavior with very weak
$T$-dependence  and absolute values
much larger than the Mott-Ioffe-Regel limit \cite{Mori98}
 indicating mean free paths much smaller than the lattice parameter.
On application of an electric field to
$\theta$-(ET)$_2$CsCo(SCN)$_4$, charge order is melted and
the conductivity is increased by several orders of magnitude
 \cite{Sawano}.

All the above 
observations suggest that an unconventional metallic
phase might be realized in such materials 
at the melting of the Wigner lattice 
driven by quantum fluctuations,
which should be further experimentally probed. Such a metal  
should display a charge gap  
  in one-electron probes 
such as photoemission or STM, or a large pseudogap
on the scale of the non-local Coulomb energy V \cite{Vlad},
together with metallic conduction   
due to the itinerancy of defects in the charge ordered background.        
The enhancement of charge fluctuations in the intermediate metallic phase 
could mediate superconductivity \cite{MM} close to the
charge ordering instability occurring, for example, in $\theta$-type
layered organic conductors and NbSe$_3$ \cite{Kiss}. 
Elucidating the precise nature of the current-carrying  
excitations emerging at zero energy in such anomalous COM phase 
remains an open challenge, as is the understanding of the interplay
between the Mott and Wigner mechanisms for the metal-insulator
transition. \cite{Camjayi08}

\section*{Acknowledgments}

S. F. acknowledges financial support from the Spanish MICINN
(Consolider CSD2007-00010) and from 
the Comunidad de Madrid through program CITECNOMIK. 
J. M. acknowledges financial support from MICINN
(project: CTQ2008-06720-C02-02).

\appendix

\section{Defect deconfinement in the Wigner lattice}
\label{defectmodel}

A system of electrons is expected to crystallize 
when the mutual Coulomb interactions dominate over the kinetic energy.
Minimizing the electrostatic interaction energy in a 
two-dimensional electron gas results in a triangular
arrangement of the charges. 
The preferred triangular ordering is altered by the presence of
an underlying periodic potential in a solid:
\cite{Uimin,Cocho,Fate}  for a density of $n=1/2$ electrons per site 
on a square host lattice the configuration of minimal 
electrostatic energy is the checkerboard order shown in Fig.\ref{fig:defects}. 
We designate such ``Wigner crystal on an underlying host lattice'' simply as
a {\it Wigner lattice} (WL).
 
As we show below, the physics of the Wigner lattice in the
strongly interacting regime $V\gg t$
can be effectively understood by considering only a small number of 
defects of the classical checkerboard, which play a dominant role in the 
low-energy excitation spectrum. These defects can be separated in two 
different classes. 
The first class corresponds to fluctuations of the charge density 
induced by quantum  fluctuations in the presence of a non-vanishing 
transfer integral $t\neq 0$, that
will be denoted as {\it neutral collective excitations}. 
Excitations of the second class, termed {\it  charged one-particle
  excitations} arise when an extra charge is added/removed
to/from the system, which is the typical situation in a photoemission
experiment. In the following paragraphs we 
analyze the two categories separately.

\subsection{Neutral, collective excitations}

In the limit of strong Coulomb interactions $V\gg t$, the excitations
of the Wigner lattice 
can be classified in terms of the potential (Madelung) energy of the
electronic configurations. 
The lowest-energy excitations of the checkerboard pattern are the
dipolar and quadrupolar defects
illustrated in Fig.\ref{fig:defects}b and c. 
These are obtained by moving respectively one or two electrons away
from their preferred equilibrium position, as shown by the arrows.  
The corresponding electrostatic energies can be evaluated to arbitrary
accuracy through standard Ewald summations, being
$\Delta_{opt}=0.6155V$ and $\Delta_{quad}=0.6453V$. The next
electronic configurations (not shown) have energies  $\ge 0.898V$ 
and will be discarded in the following discussion.

Using the procedure of Ref. \cite{Rast}, we calculate
the energy gain induced by a finite transfer integral $t$ between
molecular sites. This is achieved  by 
restoring the translational invariance and solving  the tight-binding 
problem  in Fourier
space within the reduced subspace consisting of the two 
defects described above. Accounting for  the two possible orientations of the
quadrupolar defect and  the four orientations of the
dipolar defect, we are led to diagonalise the following 
$6\times 6$ matrix (the intermolecular spacing $a$ is taken as the
unit length):
\begin{equation}
  \label{eq:matr}
  \left( \begin{array}{cccccc}
\Delta_{quad} & 0 & -t e^{ik_x}&  -t e^{-ik_x}& -t e^{ik_y}& -t e^{-ik_y}\\
0& \Delta_{quad}  & -t e^{-ik_x}&  -t e^{ik_x}& -t e^{-ik_y}& -t
e^{ik_y}\\
-t e^{-ik_x}&  -t e^{ik_x}& \Delta_{opt}&0&0&0\\
-t e^{ik_x}&  -t e^{-ik_x}& 0& \Delta_{opt}&0&0\\
-t e^{-ik_y}&  -t e^{ik_y}& 0&0& \Delta_{opt}&0\\
-t e^{ik_y}&  -t e^{-ik_y}& 0&0&0& \Delta_{opt}
 \end{array} \right)
\end{equation}
The key point here is that due to the long-range nature of the
electron-electron interactions,  the dipolar and
quadrupolar defects are almost degenerate in
energy: $\Delta_{quad}-\Delta_{opt}\simeq 0.03V$. 
As a result
a resonant tunneling is established between the two non-equivalent states 
for any physically reasonable value of the $V/t$ ratio.
The electron density is then shared between these
states, forming an ``electron droplet''.  
Such delocalisation of the charge yields a kinetic energy gain $\propto t$ and
allows the droplet to
propagate coherently at long distances with a band-like
dispersion. 
The present scenario can be contrasted to the model with
nearest-neighbour interactions, where the defects remain strongly 
localized in space unless the transfer integral overcomes  
the large energy barrier separating the states depicted in
Figs. \ref{fig:defects}b and c, i.e.  
$t\gtrsim (\Delta_{quad}-\Delta_{opt})=V$.

By setting $\Delta_{quad}=\Delta_{opt}$ in the matrix (\ref{eq:matr})
 we obtain an analytical
expression for the energy gain of the droplet state at $k=0$: 
\begin{equation}
\label{eq:gain}
  \Delta_{k=0}=\Delta_{opt} -2\sqrt{2}t.
\end{equation}
This form is in satisfactory agreement 
with the exact diagonalization (ED) result:  
$\Delta_{k=0}\simeq \Delta_{opt} - 4.2 t$,
indicating that the physical mechanism of defect delocalization
is correctly captured already in the small droplet approximation, where 
only dipolar and quadrupolar states are included.

\begin{figure}
\begin{center}
\epsfig{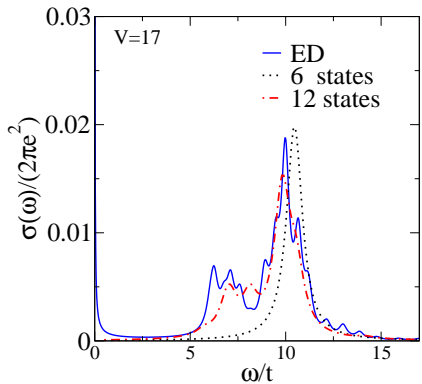}
\end{center}
\caption{(Color online) Comparison of the optical conductivity
  calculated  at $V=17t$ from ED of a $L=16$ cluster with the one obtained from
  the defect model, including respectively the 6 and 12 lowest-lying states. 
}
\label{fig:optdefect}
\end{figure}

We note that while the quadrupolar states are the ones
responsible for the band dispersion, it can be shown following the
lines of Ref. \cite{Rast} that
the optical spectral weight is mostly carried
by the dispersionless dipolar excitations. As a result, the peak in the optical
conductivity remains centered at an energy $\omega= \Delta_{opt}$
 while the  peak at $\omega= \Delta_{k=0}$ is not optically active.
The optical conductivity calculated from the defect model 
at $V=17t$ is shown in
Fig. \ref{fig:optdefect}, and compared with the spectrum obtained from
the full ED in a $L=16$ cluster.
The minimal model of Eq. (\ref{eq:matr}) 
including 6 defect states  is able to reproduce the correct position of the
exciton peak. Extending the defect subspace to a total of 12 states
also accounts for the emergence of a sideband below the main peak, as
seen in the numerical data.

\subsection{Charged, single particle excitations: interstitials and vacancies}

An analogous procedure can be carried out to determine the dispersion of
the one-electron states that are probed in  photoemission
experiments. 
It is convenient to  include a  static 
compensating charge density of $+e/2$ per site in the calculation
(equivalent to the usual
compensating jellium in continuum models), which restores the
overall charge neutrality and the particle-hole symmetry.
We can therefore focus on
the addition spectrum alone, since the removal spectrum is obtained by
symmetry.

\begin{figure}[t]
\centering
\includegraphics[width=8cm]{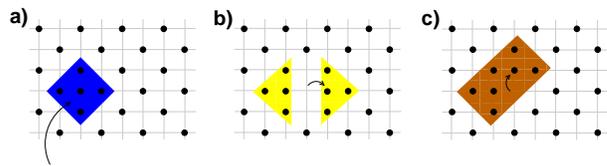}
\caption{
The lowest-lying states with one electron added to the
 perfect  checkerboard pattern. From left to right the corresponding 
Madelung energies in an infinite system 
are (a) $E_I=0.8078V$, (b) $E_D=0.9233$ and (c) 
$E_M = 1.1304$. }
  \label{fig:interst}
\end{figure}

As usual we start from the $V\gg t$ limit. 
Adding  an electron to an empty site of the Wigner lattice
creates an interstitial, of energy $E_I=0.8078V$. 
The next low-lying states with $N+1$ electrons have energies respectively 
 $E_D=0.9233$  and $E_M = 1.1304$. All these states are illustrated in Fig.\ref{fig:interst}. 
The charge gap is defined  as $\Delta_{ch}=E(N+1)+E(N-1)-2E(N)$ which gives 
$\Delta_{ch}=2|E_I|=1.6155V$. 

In order to study the effect of quantum fluctuations induced by a finite
transfer integral $t$ we  evaluate the
action of the  kinetic energy operator within the subspace of the
defect states defined above.   
Considering the multiplicities arising from the possible
orientations of these states (respectively $2,4,8$)   
one obtains the  $14\times 14$ matrix shown in Table \ref{tab:14matr}.
The charge gap at finite $t$ 
is obtained from  the energy of the lowest state at $k=0$. In the
region $10<V/t<20$ it follows the linear form  
$\Delta_{ch}\simeq 1.8 V- 9.6 t$, in good agreement
with the ED value: $\Delta_{ch}\simeq 1.8 V- 7.0 t$.

To get more insight on the one-particle excitations of the Wigner lattice, we 
observe that the  bands  arising from the
diagonalization of the matrix in Table \ref{tab:14matr} are 
accurately described by
the following formula: 
\begin{eqnarray}
  \label{eq:disp}
E_k &=& E_0-2 t_1 [\cos (2 k_x) + \cos (2 k_y)] \\ 
\nonumber & & -
2 t_2 [\cos (k_x + k_y) + \cos(k_x - k_y)].
\end{eqnarray}
An important implication of
the defect model is that the  hopping parameters   
$t_1,t_2$ governing the motion of defects 
are essentially constant throughout
the charge ordered phase: the one-electron 
band dispersion depends on the interaction strength $V$ only via the
rigid energy shift $E_0$. 
For a direct validation of  the defect model, 
Fig.\ref{fig:spectral} shows how the  spectral function $A(k,\omega)$ obtained
from the exact diagonalisation of a $L=16$ cluster within the charge
ordered metallic phase closely follows the dispersion of the main defect band 
[Eq. (\ref{eq:disp}), with  $t_1=0$, $t_2= -0.3 t$], which represents the
motion of interstitials along the diagonals of the square
lattice.

\section{Finite-size effects}
As was pointed out in Ref.\cite{Tsiper}, in addition to the usual
size effects encountered in numerical studies of model
systems on finite clusters (hereafter denoted as {\it quantum finite-size
effects}), specific issues arise when dealing 
with the  long-range  Coulomb potential. These originate from the
inaccuracy of calculating the 
electrostatic energy of the electronic configurations 
in a finite system. Such effects are termed
{\it classical finite size effects} and become relevant within the
charge ordered phase at large $V/t$.

\subsection{Quantum finite-size effects}

Ordinary finite size effects in the EHM have been explored
previously by comparing results of different quantities with different number of cluster
sites $L=8,10,16$ and $20$\cite{Calandra02}. 
Other possible ways of evaluating finite size effects are based on averaging
over the boundary condition \cite{Koretsune}. In the case of the extended Hubbard model
both procedures are found to lead to similar critical values\cite{Seo}.

We have performed finite-size scaling for the HWM by evaluating the
charge correlation function, $C(\pi,\pi)$, Drude weight, $D$, and
charge gap, $\Delta_{ch}$ on $L=8,10,16,18$ clusters, and analyzing
the extrapolation to the thermodynamic limit. In Fig. \ref{fig:scaling}
we show $C(\pi,\pi)$ plotted as a function of $1/L$ and the Drude weight, $D$,
dependence with $1/\sqrt{L}$. The finite size 
scaling of the single particle charge gap, $\Delta_{ch}$, is shown in
Fig. \ref{fig:chargegap}. 
Our finite-size scaling analysis strongly suggests
the presence of a broad metallic phase with both charge order and a finite 
single-particle  charge gap for $V_{CO}<V<V_{MI}$.

\subsection{Classical finite-size effects}

Classical finite-size effects are minimized by evaluating
electrostatic energies with Ewald summations, which amounts to performing
an infinite periodic repetition of the finite simulation cell. Although a
system of infinite size is effectively recovered through this
procedure, this gives rise to
spurious contributions to the energy arising from the interaction of
a given electronic configuration with its images on the repeated cells.
Since the simulation cell is overall neutral, the
interaction between equivalent cells 
has a dipolar nature and the corresponding  error in the energy
scales at most as $\Delta E\sim V/L^{3/2}$ ($L$ being the total number
of sites). 

Having identified in Appendix \ref{defectmodel}
 those electronic configurations that are mainly 
responsible for the low-energy behaviour of the Wigner lattice, we can 
precisely address the effect of such uncertainties on the
physics of interest here. To this aim we report  in  Table \ref{tab:energies}  
the electrostatic 
energies of the configurations used in the defect model of 
Appendix \ref{defectmodel}, and compare them  with the corresponding
values in a $4\times 4$ cluster (energies are expressed in units of
$V$). 
As we can see, finite-size 
errors on these states are typically of the order of $5-10\%$ or less,
which gives us good confidence on the final numerical results. \\

\begin{table}[h]
  \centering
\[\begin{array}{|c|ccccc|}
    \hline  &  E_{opt}& E_{quad} & E_I&E_D&E_M\\
\hline extended & 0.6155&  0.6453& 0.8078 & 0.9233 & 1.1304\\
4\times 4 & 0.5776& 0.6258 &0.8078& 1.0370&1.1207 \\ 
\hline
\end{array}\]
  \caption{Electrostatic energies of the relevant defect states.}
  \label{tab:energies}
\end{table}

\pagebreak

\begin{sidewaystable}
\label{tab:14matr}
\caption{The $14\times 14$ matrix used to calculate the one-particle
  excitation spectrum in the defect subspace.}
{\footnotesize
\[\left( 
\begin{array}{cccccccccccccc}
E_I & 0 & -2t \cos k_x & -2t \cos k_y&0&0 &
 -t e^{i\frac{k_x+k_y}{2}}& -t e^{i\frac{k_x+k_y}{2}}& 
 -t e^{-i\frac{k_x+k_y}{2}}& -t e^{-i\frac{k_x+k_y}{2}}& 
 -t e^{-i\frac{k_x-k_y}{2}}& -t e^{-i\frac{k_x-k_y}{2}}& 
 -t e^{i\frac{k_x-k_y}{2}}& -t e^{i\frac{k_x-k_y}{2}} \\
0& E_I  & -2t \cos k_x & -2t \cos k_y&0&0 &
 -t e^{-i\frac{k_x+k_y}{2}}& -t e^{-i\frac{k_x+k_y}{2}}& 
 -t e^{i\frac{k_x+k_y}{2}}& -t e^{i\frac{k_x+k_y}{2}}& 
 -t e^{i\frac{k_x-k_y}{2}}& -t e^{i\frac{k_x-k_y}{2}}& 
 -t e^{-i\frac{k_x-k_y}{2}}& -t e^{-i\frac{k_x-k_y}{2}} \\
-2t \cos k_x & 0& E_D&0&0&0& 0&0&0&0&0&0&0&0\\
-2t \cos k_y & 0& 0&E_D&0&0& 0&0&0&0&0&0&0&0\\
0&-2t \cos k_x & 0&0& E_D&0&0&0& 0&0&0&0&0&0\\
0&-2t \cos k_y & 0& 0& 0&E_D&0&0& 0&0&0&0&0&0\\
-t e^{-i\frac{k_x+k_y}{2}}& -t e^{i\frac{k_x+k_y}{2}}&
0&0&0&0&E_M&0&0&0&0&0&0&0\\
-t e^{-i\frac{k_x+k_y}{2}}& -t e^{i\frac{k_x+k_y}{2}}&
0&0&0&0&0&E_M&0&0&0&0&0&0\\
-t e^{i\frac{k_x+k_y}{2}}& -t e^{-i\frac{k_x+k_y}{2}}&
0&0&0&0&0&0&E_M&0&0&0&0&0\\
-t e^{i\frac{k_x+k_y}{2}}& -t e^{-i\frac{k_x+k_y}{2}}&
0&0&0&0&0&0&0&E_M&0&0&0&0\\
-t e^{i\frac{k_x-k_y}{2}}& -t e^{-i\frac{k_x-k_y}{2}}&
0&0&0&0&0&0&0&0&E_M&0&0&0\\
-t e^{i\frac{k_x-k_y}{2}}& -t e^{-i\frac{k_x-k_y}{2}}&
0&0&0&0&0&0&0&0&0&E_M&0&0\\
-t e^{-i\frac{k_x-k_y}{2}}& -t e^{i\frac{k_x-k_y}{2}}&
0&0&0&0&0&0&0&0&0&0&E_M&0\\
-t e^{-i\frac{k_x-k_y}{2}}& -t e^{i\frac{k_x-k_y}{2}}&
0&0&0&0&0&0&0&0&0&0&0&E_M
 \end{array} 
\right)\] }
\end{sidewaystable}

\end{document}